\documentclass[sigconf, nonacm]{acmart}

\AtBeginDocument{%
  }

\usepackage{multirow}
\usepackage{booktabs}
\usepackage{graphicx}
\usepackage{hyperref}
\usepackage{tikz}
\usetikzlibrary{positioning, arrows.meta, shapes.geometric}
\usetikzlibrary{calc}

\usepackage{balance}

\begin{document}

\title{Internet-Mediated Digital Informal Learning Portfolios in STEM Higher Education: A Computational Grounded Theory Study of Online Peer Advice Communities}

\author{Jianjun Xiao}
\email{et\_shaw@126.com}
\orcid{0000-0003-0000-9630}
\affiliation{%
  \institution{Faculty Of Education, Beijing Normal University}
  \city{Beijing}
  % \state{}
  \country{China}
}

\author{Yuxi Long}
\authornote{Corresponding author.}
\email{202331010035@mail.bnu.edu.cn}
\orcid{0009-0007-0436-6797}
\affiliation{%
  \institution{Faculty Of Education, Beijing Normal University}
  \city{Beijing}
  \country{China}}

\renewcommand{\shortauthors}{Xiao et al.}
\renewcommand{\shorttitle}{Internet-Mediated Digital Informal Learning Portfolios}

\begin{abstract}
Internet technologies have expanded higher education students' access to learning
resources, peer guidance, and skill-development opportunities beyond formal curricula.
Yet the ways students assemble these distributed online resources into coherent
learning pathways remain insufficiently understood. This study examines how STEM
students construct digital informal learning portfolios through internet-mediated
peer advice and platform use. Drawing on Social Cognitive Career Theory (SCCT) and
informal learning frameworks, we analyze 3,607 peer advice posts from a large online
student community using Computational Grounded Theory (CGT). Results show that career pathway (69.6\% of coded documents) and career
orientation (59.7\%) are the dominant organizing dimensions, yielding three distinct
digital informal learning portfolios: a \textit{graduate-study portfolio} centered on
competition training, mathematical foundations, and staged preparation; an
\textit{industry-employment portfolio} centered on self-directed skill building,
online platform learning, and strategically timed internships; and a \textit{public-sector portfolio} characterized
by dual-track hedging across graduate study, enterprise employment, and public-sector preparation pathways. The online peer
community itself functions as a distributed informal curriculum, collectively producing
and transmitting pathway-specific guidance about what to learn, when to learn it, and
which internet resources to prioritize. These findings extend SCCT into the domain of
internet-mediated digital informal learning and introduce \textit{career front-loading}
as a pattern of early learning reorganization. Implications are
discussed for institutional learning support, recognition of internet-enabled learning,
and the design of digital guidance infrastructures in higher education.
\end{abstract}

\keywords{Higher education; Digital informal learning; Online peer communities; Computational grounded theory; Digital learning portfolios}

\maketitle

%%
%% Main content.
\section{Introduction}\label{sec:introduction}

The internet has become an integral part of higher education students' learning environments, expanding access to resources, peer guidance, and skill-development opportunities beyond formal curricula \citep{dabbagh2012personal, littlejohn2016learning}. In science, technology, engineering, and mathematics (STEM) fields in particular, students increasingly build learning trajectories across formal coursework and internet-mediated informal spaces. Yet we still know relatively little about how they combine these dispersed opportunities into coherent, goal-oriented patterns of learning.

A growing body of research shows that students do not rely on institutional provision alone, but actively pursue self-directed informal learning through multiple digital platforms \citep{livingstone1999exploring, colley2003informality}. Within this broader ecology, online peer communities have become especially consequential. They do not merely provide isolated tips; they also circulate pathway-specific interpretations of what is worth learning, where to learn it, and how learning activities should be sequenced in relation to anticipated futures.

However, three limitations remain in the existing literature. First, prior studies often focus on single platforms, single practices, or broad notions of self-directed learning, leaving unclear how students assemble multiple internet-enabled learning activities into larger configurations. Second, research on informal learning in higher education has paid limited attention to how anticipated career pathways shape these configurations, despite the relevance of Social Cognitive Career Theory \citep[SCCT;][]{lent1994toward}. Third, studies of online peer communities have typically treated them as support spaces or information sources rather than as archives through which collective guidance about learning pathways can be theorized.

To address these gaps, this study analyzes 3,607 peer advice posts from a large Chinese online student community spanning 38 STEM majors, a setting in which major-specific career competition is salient and peer advice is extensively produced, circulated, and publicly archived. Using Computational Grounded Theory \citep[CGT;][]{nelson2020computational}, we identify recurring patterns in how pathway-specific advice organizes digital informal learning. In doing so, we develop the concepts of the \textit{digital informal learning portfolio} and \textit{career front-loading} to explain how extracurricular learning investments are recommended, prioritized, and sequenced early in the undergraduate years.

This study makes three contributions to the literature on internet-mediated learning in higher education. First, it introduces the concept of the \textit{digital informal learning portfolio} as an analytic lens for understanding how students combine and sequence multiple online learning activities in relation to anticipated career pathways. Second, it shows that peer advice frames career expectations as persistent portfolio-organizing forces, thereby extending SCCT into the analysis of digital informal learning, and introduces \textit{career front-loading} as a concept for understanding the early career-oriented framing of these learning investments. Third, it conceptualizes the online peer advice community as a distributed informal curriculum that transmits pathway-specific guidance about valued skills, appropriate timing, and the conversion of informal learning into recognizable educational and career advantages.

The remainder of this paper is organized as follows. Section~\ref{sec:theoretical_background} presents the theoretical background, situating the study within SCCT and the informal learning literature. Section~\ref{sec:literature_review} reviews related work on digital informal learning platforms and peer advice communities. Section~\ref{sec:methodology} describes the CGT methodology, including data collection, computational text analysis, and qualitative coding procedures. Section~\ref{sec:results} reports the findings, presenting the three portfolio types and their constituent learning activities. Section~\ref{sec:discussion} discusses the theoretical and practical implications of the results. Section~\ref{sec:conclusion} concludes with a summary of contributions, limitations, and directions for future research.

\section{Theoretical Background}\label{sec:theoretical_background}

\subsection{Informal Learning in Digital Higher Education Contexts}

The concept of informal learning predates the digital era but has acquired renewed salience in contemporary higher education as internet technologies increasingly mediate how students access knowledge, guidance, and learning opportunities. \citet{livingstone1999exploring} defines informal learning as the pursuit of understanding, knowledge, or skill outside the curricula of formal institutions. This definition foregrounds two core attributes: informal learning is self-initiated by the learner, and it occurs beyond formally structured educational settings.

\citet{colley2003informality} refine this concept by arguing that formality and informality should be understood as a continuum rather than a binary opposition. This perspective is particularly useful for analyzing internet-enabled learning activities that remain self-directed and extra-institutional even when they borrow structured features from formal education. \citet{marsick2001informal} further distinguish intentional informal learning from incidental learning, underscoring that not all learning outside institutions is organized in the same way.

What remains undertheorized in this literature is how students combine multiple internet-based informal learning channels in ways that reflect career-oriented learning patterns. A student preparing for graduate school may assemble a very different set of platforms and activities than one targeting enterprise employment, yet existing research tends to examine individual platforms in isolation. This paper introduces the concept of a \textit{digital informal learning portfolio}---the career-directed combination of multiple digital informal learning channels that a student assembles and sequences over time. Throughout this paper, the term \textit{career-directed} modifies the portfolio concept and its analytical framework, denoting purposive orientation toward anticipated career outcomes; \textit{career-oriented} is used descriptively to characterize the discourse patterns and student behaviors observed in the data. This concept draws on the informal learning tradition \citep{livingstone1999exploring, colley2003informality} while foregrounding the strategic, multi-platform nature of contemporary higher education learning in internet-mediated environments.

\subsection{Social Cognitive Career Theory and Learning Behavior}

Understanding why students construct particular informal learning portfolios requires a theoretical account of how career aspirations shape learning decisions. SCCT, developed by \citet{lent1994toward} and grounded in social cognitive theory, provides such an account. The framework emphasizes self-efficacy, outcome expectations, and personal goals as the key mechanisms through which individuals organize career-related behavior.

\citet{lent1994toward} argue that these variables jointly shape career interest formation, career choice, and performance attainment. Of these, outcome expectations are especially relevant here because they help explain why students invest in some learning activities while neglecting others. \citet{lent2002social} subsequently extended SCCT to academic persistence and performance, showing that the same mechanisms that shape career choice can also organize ongoing learning behavior. Students who hold strong outcome expectations for a particular pathway are therefore likely to invest disproportionately in activities aligned with that pathway.

The mechanism through which career outcome expectations translate into organized learning behavior can be further clarified through self-regulation theory. \citet{zimmerman2000attaining} conceptualize self-regulated learning as a cyclical process of forethought, performance, and self-reflection. In this study, self-regulation serves as a bridge between distal career goals and proximal learning decisions.

Existing SCCT research has examined how self-efficacy and outcome expectations predict career interests, choice actions, and persistence in academic programs \citep{lent1994toward, lent2002social}. Recent work in STEM educational contexts similarly suggests that anticipated career destinations structure how students engage with learning opportunities \citep{jiang2024influence, markman2024career}. What remains unresolved is how career outcome expectations organize the composition and sequencing of digital informal learning activities across multiple platforms.

\subsection{Analytical Framework}

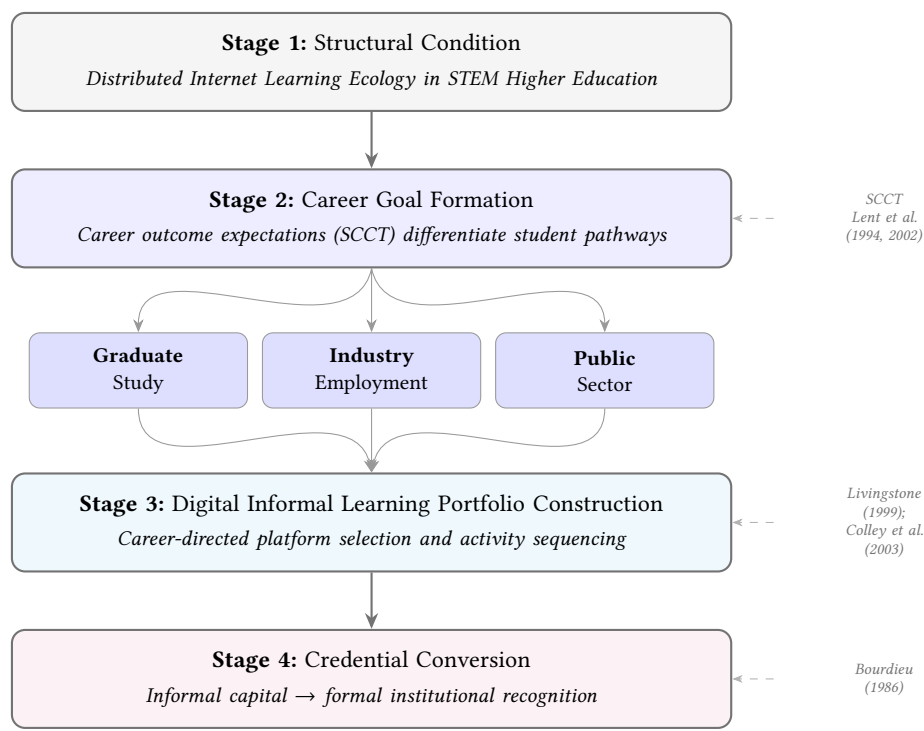
\begin{figure*}[ht]
\centering
\begin{tikzpicture}[
  node distance = 0.75cm,
  %% Box styles
  stagebox/.style = {
    rectangle, draw=black!55, thick, rounded corners=5pt,
    text width=9cm, minimum height=1.15cm,
    align=center, inner sep=8pt
  },
  pathbox/.style = {
    rectangle, draw=black!40, rounded corners=4pt,
    text width=2.55cm, minimum height=1.0cm,
    align=center, inner sep=5pt, font=\small
  },
  annonode/.style = {
    font=\scriptsize\itshape, text=black!55,
    text width=2.7cm, align=center
  },
  %% Arrow styles
  arr/.style   = {-{Stealth[length=6pt,width=5pt]}, thick, black!55},
  subarr/.style= {-{Stealth[length=5pt,width=4pt]}, black!40},
  dann/.style  = {-{Stealth[length=4pt,width=3pt]}, dashed, black!30},
]

%% ── Stage 1 ──────────────────────────────────────────────
\node[stagebox, fill=gray!8] (S1) {
  \textbf{Stage 1:}~Structural Condition\\[2pt]
  {\small\itshape Distributed Internet Learning Ecology in STEM Higher Education}
};

%% ── Stage 2 ──────────────────────────────────────────────
\node[stagebox, fill=blue!7, below=0.75cm of S1] (S2) {
  \textbf{Stage 2:}~Career Goal Formation\\[2pt]
  {\small\itshape Career outcome expectations (SCCT) differentiate student pathways}
};

%% ── Three career pathways ────────────────────────────────
\node[pathbox, fill=blue!13, below=0.85cm of S2, xshift=-3.1cm] (P1) {
  \textbf{Graduate}\\Study
};
\node[pathbox, fill=blue!13, below=0.85cm of S2] (P2) {
  \textbf{Industry}\\Employment
};
\node[pathbox, fill=blue!13, below=0.85cm of S2, xshift= 3.1cm] (P3) {
  \textbf{Public}\\Sector
};

%% ── Stage 3 ──────────────────────────────────────────────
\node[stagebox, fill=cyan!6, below=0.85cm of P2] (S3) {
  \textbf{Stage 3:}~Digital Informal Learning Portfolio Construction\\[2pt]
  {\small\itshape Career-directed platform selection and activity sequencing}
};

%% ── Stage 4 ──────────────────────────────────────────────
\node[stagebox, fill=purple!5, below=0.75cm of S3] (S4) {
  \textbf{Stage 4:}~Credential Conversion\\[2pt]
  {\small\itshape Informal capital~$\rightarrow$~formal institutional recognition}
};

%% ── Main vertical arrows ─────────────────────────────────
\draw[arr] (S1) -- (S2);
\draw[arr] (S3) -- (S4);

%% ── S2 fans out to three pathways ────────────────────────
\draw[subarr] (S2.south) to[out=258,in=90] (P1.north);
\draw[subarr] (S2.south)                -- (P2.north);
\draw[subarr] (S2.south) to[out=282,in=90] (P3.north);

%% ── Three pathways converge into S3 ─────────────────────
\draw[subarr] (P1.south) to[out=270,in=102] (S3.north);
\draw[subarr] (P2.south)                 -- (S3.north);
\draw[subarr] (P3.south) to[out=270,in= 78] (S3.north);

%% ── Theory annotations (right side) ─────────────────────
\node[annonode, right=0.55cm of S2] (A2) {SCCT\\Lent et al.\\(1994, 2002)};
\node[annonode, right=0.55cm of S3] (A3) {Livingstone\\(1999);\\Colley et al.\\(2003)};
\node[annonode, right=0.55cm of S4] (A4) {Bourdieu\\(1986)};

\draw[dann] (A2.west) -- (S2.east);
\draw[dann] (A3.west) -- (S3.east);
\draw[dann] (A4.west) -- (S4.east);

\end{tikzpicture}
\caption{Career-directed digital informal learning portfolio framework.
  Dashed arrows indicate the theoretical grounding of each stage.}
\label{fig:framework}
\end{figure*}

Drawing on the theoretical foundations outlined above, this study proposes a four-stage analytical framework (Figure~\ref{fig:framework}) that traces the pathway from structural conditions to individual learning behavior.

\textbf{Stage 1: Distributed internet learning ecology as structural condition.} The framework begins with the structural context of contemporary STEM higher education: important learning resources, guidance, and skill-development opportunities are distributed across both formal curricula and internet-based environments. This distribution creates a gap between what formal programs provide and what students perceive they need for career success, motivating students to seek supplementary learning opportunities online.

\textbf{Stage 2: Career goal formation mediated by SCCT outcome expectations.} Faced with this gap, students do not respond uniformly. Drawing on SCCT, we argue that students' responses are organized by career outcome expectations---beliefs about the likely consequences of pursuing particular career pathways. These expectations differentiate students into distinct pathway orientations (graduate study, enterprise employment, public-sector preparation), each associated with different capability requirements and timeline pressures.

\textbf{Stage 3: Digital informal learning portfolio construction.} Once pathway orientations are established, they function as organizing principles for learning investment decisions. Students select and sequence digital informal learning activities---MOOCs, competitions, online practice platforms, internships---in ways aligned with their chosen pathway's capability demands and temporal structure. The resulting portfolio represents a coherent, career-directed configuration of informal learning investments that extends beyond formal curriculum boundaries.

\textbf{Stage 4: Credential conversion.} The final stage addresses how informal learning outputs acquire value within formal institutional contexts. Competitions, projects, and self-directed learning achievements function as forms of capital, but their value depends on conversion into institutionally recognized credentials---admissions offers, employment opportunities, or certification. This conversion process determines whether informal learning investments translate into tangible career advantages.

This framework raises empirical questions about portfolio structure, career-orientation differentiation, and construction mechanisms---questions we address through CGT analysis in the following sections.

\section{Literature Review}\label{sec:literature_review}

\subsection{Internet-Mediated Learning Gaps in STEM Higher Education}

A substantial body of research documents persistent gaps between what STEM higher education curricula provide and what students perceive they need for future study and work. In internet-rich learning environments, these gaps matter not only because formal curricula may be incomplete, but because students can now turn to a wide array of online resources and communities to compensate for perceived shortcomings. \citet{crawley2014rethinking} articulate this problem through the CDIO initiative, an international engineering education reform framework premised on the observation that traditional curricula overemphasize decontextualized disciplinary knowledge while underdelivering on the integrative competencies---design thinking, teamwork, system-level reasoning---that professional practice requires. The CDIO framework represents one of the most comprehensive institutional responses to this mismatch, yet its proponents acknowledge that curricular reform alone cannot fully close the gap between educational provision and professional demand.

The employability literature provides a complementary perspective. \citet{clarke2018rethinking} argues that graduate employability cannot be reduced to credential acquisition; rather, it encompasses a complex assemblage of disciplinary knowledge, transferable skills, and career management competencies that formal degree programs address unevenly. Clarke's analysis reveals a structural tension: employers increasingly evaluate candidates on demonstrated capabilities and practical experience, while higher education institutions continue to certify graduates primarily through examination-based assessment of theoretical knowledge. This credential--competence gap is not merely a matter of curricular content but reflects deeper institutional logics governing how universities define and measure student achievement.

In computer science and software engineering, the mismatch takes particularly concrete forms. \citet{garousi2020closing} conduct a systematic mapping study comparing topics covered in software engineering curricula with those prioritized by industry practitioners, finding significant discrepancies in areas such as software testing, configuration management, and agile development practices. Their analysis demonstrates that curricular revision, while necessary, operates on institutional timescales that cannot match the pace of technological change in the software industry. The resulting gap creates a persistent structural condition that students must navigate throughout their undergraduate careers. More recent empirical work confirms the persistence and breadth of this condition: \citet{akdur2023skills} documents systematic mismatches between industry-required and university-developed skills across multiple software engineering roles, particularly in configuration management, software quality assurance, and agile practices, while \citet{dobslaw2023gap} trace technology-level divergences between institutional syllabi and industry job postings, finding that cloud and automation technologies are heavily demanded by employers yet remain peripheral in most academic curricula.

A critical observation about this body of literature is that it overwhelmingly focuses on institutional responses rather than on how students themselves navigate these gaps through internet-enabled learning and peer guidance. This leaves student-level adaptive strategy comparatively underexamined.

\subsection{Digital Informal Learning in Higher Education}

A growing body of research examines the digital platforms and environments through which higher education students pursue learning outside formal curricula. This literature spans multiple platform types, each studied largely in isolation.

MOOCs have attracted the most sustained scholarly attention as informal learning vehicles. \citet{littlejohn2016learning} investigate how learners engage with MOOCs and find that self-regulation capacity significantly differentiates learning outcomes. Crucially, their research reveals that many MOOC participants are not formal enrollees seeking certification but self-directed learners who selectively access course materials---lectures, readings, problem sets---to address specific knowledge gaps \citep{zhang_learning_2021}. This instrumental, goal-directed pattern of MOOC use aligns more closely with informal learning as defined by \citet{livingstone1999exploring} than with the formal enrollment model that MOOC platforms nominally instantiate.

Online judge systems and algorithmic training platforms constitute another significant category of digital informal learning environments, particularly in computer science education. \citet{wasik2018survey} provide a comprehensive survey of these systems, documenting their pedagogical architecture: structured problem sets of graduated difficulty, automated evaluation through test cases, competitive ranking mechanisms, and community-generated editorial content. More recent work by \citet{wang2023problem} demonstrates that problem difficulty and sequencing in online judge systems exert measurable effects on students' learning outcomes and cognitive load, providing evidence that platform design choices constitute meaningful pedagogical decisions with real consequences for skill acquisition. These platforms occupy an interesting position on the formality--informality continuum described by \citet{colley2003informality}: their content is highly structured and their assessment mechanisms are rigorous, yet participation is entirely voluntary, self-paced, and unconnected to institutional credit systems.

Open-source software communities, particularly those organized around GitHub, represent a third category of informal learning environment. \citet{feliciano2016student} examine how students use GitHub in the context of software engineering courses and find that platform participation exposes students to professional development practices---version control, code review, collaborative workflows---that formal curricula often reference but rarely operationalize at scale. GitHub participation thus provides a form of situated learning \citep{lave1991situated} in which students acquire professional competencies through legitimate peripheral participation in authentic practice communities.

Research on self-regulated learning in online higher education contexts provides insight into the cognitive and metacognitive processes that underpin digital informal learning. \citet{broadbent2015selfregulated} conduct a systematic review of self-regulation strategies in online higher education and identify time management, metacognition, effort regulation, and critical thinking as strategies positively associated with academic outcomes. Their findings suggest that successful digital informal learning requires sophisticated self-regulatory capacity, yet the review focuses on formal online courses rather than the voluntary, self-initiated learning activities that constitute informal learning portfolios. \citet{dabbagh2012personal} advance the concept of personal learning environments (PLEs), arguing that students use social media and Web 2.0 tools to create self-directed learning spaces that support goal setting, information management, and social interaction. The PLE concept captures the multi-platform nature of contemporary student learning but does not address how career orientations shape PLE composition. Examining this dynamic in a Chinese higher education context, \citet{liu2025bridging} find that students engage with digital informal learning primarily to compensate for perceived gaps between formal curricula and career aspirations, constructing their digital learning activities around possible-self visions of academic and professional success---a motivation structure that resonates with the career-directed portfolio logic this study proposes.

Two limitations characterize this literature. First, studies overwhelmingly examine single platforms or platform types in isolation rather than integrated, portfolio-level configurations. Second, the role of career goals in organizing digital informal learning remains undertheorized.

\subsection{Online Peer Communities and Knowledge Sharing in Higher Education}

Beyond formal and informal learning platforms, online peer communities constitute a distinctive source of learning resources for higher education students. The theoretical foundation for understanding these communities derives from \citeauthor{lave1991situated}'s (\citeyear{lave1991situated}) concept of CoP, originally developed to explain how learning occurs through participation in social groups organized around shared domains of interest. In the digital era, this concept extends naturally to online forums, social media groups, and question-and-answer platforms where students exchange knowledge, experience, and strategic advice.

Online peer advice communities differ from both formal instruction and self-directed platform learning in a theoretically significant way: the knowledge they circulate is generated by peers who occupy similar structural positions rather than by instructors or platform designers who occupy positions of institutional authority \citep{anderson_three_2011}. This peer-generated knowledge carries distinctive epistemic properties \citep{chen_new_2025}. It reflects the lived experience of navigating educational systems, internet-based learning resources, labor markets, and career transitions from the student's perspective. It encodes tacit knowledge about effective learning strategies, platform selection, resource sequencing, and credential conversion that institutional actors may not possess or may not communicate. In this sense, online peer communities function as informal curriculum providers \citep{xu_connectivist_2024}, offering experientially grounded guidance that complements---and sometimes contradicts---official institutional advice.

Despite their evident influence on student learning behavior, online peer advice communities remain understudied as primary data sources for theory building in higher education research. Existing studies typically treat them as sites for investigating predefined variables rather than as corpora from which new constructs can be inductively derived. In particular, CGT has not been applied to online peer advice data in this domain.

\subsection{Computational Grounded Theory}

Grounded theory (GT) provides a systematic methodology for generating theory from empirical data through iterative coding and constant comparison \citep{glaser1967discovery}. Rather than testing pre-specified hypotheses, GT researchers inductively develop conceptual categories and theoretical propositions that are ``grounded'' in the data themselves. Within the GT tradition, the constructivist variant advanced by \citet{charmaz2006constructing} foregrounds researcher reflexivity and acknowledges that theoretical insights are co-constructed through the analyst's engagement with participants' meanings. This epistemological stance is well suited to studying online peer advice, where students articulate subjective learning experiences and career aspirations in naturalistic settings.

A persistent limitation of traditional GT, however, is its reliance on small, purposively sampled datasets. When the phenomenon of interest is distributed across thousands of user-generated texts, manual line-by-line coding becomes impractical, yet purely automated approaches such as unsupervised topic modeling sacrifice the theoretical sensitivity that distinguishes GT from generic content analysis.

CGT, proposed by \citet{nelson2020computational}, addresses this tension by integrating computational text-analysis techniques into each phase of the GT workflow while preserving human interpretive judgment at every decision point. In CGT, computational methods serve as a ``scaffolding'' that extends the analyst's reach without replacing theoretical reasoning. Pattern-detection algorithms assist with open coding, statistical co-occurrence measures inform axial coding, and network-analytic metrics guide selective coding---yet the analyst retains authority over codebook construction, category naming, and theory articulation. A subsequent methodological refinement by \citet{carlsen2022computational} argues for computer-\textit{assisted} rather than computer-\textit{led} analysis, cautioning that unsupervised algorithms should not be expected to produce semantically coherent categories autonomously; instead, human interpretive engagement must actively direct and evaluate each computational step to ensure that emergent constructs are grounded in substantive meaning rather than algorithmic artifact.

\subsection{Research Gap}

Taken together, these literatures leave three linked gaps: the absence of portfolio-level accounts of internet-mediated informal learning, the underuse of online peer advice data for theory building, and the lack of empirical work linking career outcome expectations to informal learning portfolio construction in higher education internet environments. These gaps motivate the research questions that follow:

\begin{itemize}
    \item[\textbf{RQ1.}] What patterns of digital informal learning behavior do STEM students report in online peer advice communities as they navigate distributed learning opportunities beyond formal curricula?
    \item[\textbf{RQ2.}] How do students' career orientations shape the composition of their digital informal learning portfolios?
\end{itemize}
\section{Methodology}\label{sec:methodology}

\subsection{Methodological Rationale}

The present study transfers the CGT framework from its original sociological application to newspaper corpora \citep{nelson2020computational} into higher education research on online peer communities. Two considerations motivate this transfer. First, the research objective is exploratory: we aim to generate a mid-range theory rather than test existing models. Second, the corpus of 3,607 peer advice posts occupies a methodological middle ground---too large for purely manual coding, yet too small and domain-specific for unsupervised machine-learning algorithms to produce semantically coherent clusters without extensive human guidance. CGT offers a principled approach for corpora of this scale, combining computational efficiency with interpretive rigor.

\subsection{Data: Online Peer Advice Community}

The data for this study are drawn from a large Chinese online student community (https://www.kkdaxue.com/). The community is a dedicated ``major advice'' forum where upper-year students and recent graduates voluntarily share experiential knowledge about their fields of study, including course recommendations, skill-building strategies, career pathways, and reflections on academic decision-making. These posts constitute a form of naturalistic, non-reactive data: they are produced spontaneously by community members for peer audiences, without researcher solicitation or intervention, thereby minimizing social desirability bias and demand characteristics that can affect interview or survey responses \citep{livingstone1999exploring}.

The raw dataset comprised 10,053 advice posts spanning all disciplines. To focus the analysis on STEM higher education, we applied three filtering criteria sequentially. First, posts were retained only if their declared major matched one of 38 pre-specified STEM discipline keywords (e.g., Computer Science, Software Engineering, Electrical Engineering, Mechanical Engineering, Civil Engineering, Automation). Second, posts shorter than 20 characters were excluded, as they typically contained only a major name or a brief greeting without substantive advice content. Third, posts with platform moderation statuses of ``deleted'' or ``review failed'' were removed. After filtering, the final analytical sample contained 3,607 posts. Table~\ref{tab:sample} summarizes the sample characteristics.

\begin{table}[h]
\centering
\caption{Sample characteristics of the 3,607 analyzed posts}
\begin{tabular}{llr}
\toprule
\textbf{Variable} & \textbf{Category} & \textbf{n (\%)} \\
\midrule
Education level & Undergraduate & 2,615 (72.5\%) \\
                & Community college & 525 (14.5\%) \\
                & Master's & 410 (11.4\%) \\
                & Doctoral & 57 (1.6\%) \\
\midrule
Year posted     & 2022 & 2,358 (65.4\%) \\
                & 2023 & 767 (21.3\%) \\
                & 2024 & 268 (7.4\%) \\
                & 2025 & 185 (5.1\%) \\
                & 2026 & 29 (0.8\%) \\
\midrule
Top 3 majors    & Computer Science & 380 (10.5\%) \\
                & Software Engineering & 252 (7.0\%) \\
                & Civil Engineering & 150 (4.2\%) \\
\midrule
\multirow{2}{*}{Text length} & Mean (SD) & 294.3 ($\pm$357.8) chars \\
                & Median & 175 chars \\
\bottomrule
\end{tabular}
\label{tab:sample}
\end{table}

The corpus is predominantly composed of undergraduate posts (72.5\%), consistent with the platform's user base, with smaller contributions from community college students (14.5\%), master's students (11.4\%), and doctoral students (1.6\%). Posts were collected from 2022 through early 2026, with the majority (65.4\%) originating in 2022. The substantial variation in text length (median 175 characters, mean 294.3 characters) indicates that the corpus encompasses both concise bullet-point recommendations and extended narrative reflections, providing rich material for theory generation.

Regarding research ethics, the data are publicly accessible online content that users anonymously posted voluntarily to a public forum. Findings are reported at the aggregate group level; no individual case analyses are presented. The research team obtained an institutional ethics review exemption for the use of this publicly available, de-identified dataset.

\subsection{Three-Phase CGT Analysis}

Following the CGT framework \citep{nelson2020computational,carlsen2022computational}, the analysis proceeded through three phases: open coding, axial coding, and selective coding. Each phase combined computational procedures with human interpretive judgment.

\subsubsection{Phase 1: Open Coding}

The open coding phase aimed to develop a comprehensive codebook and apply it systematically to the full corpus. Two researchers independently read a stratified random sample of 200 posts, writing analytic memos to capture emergent themes related to learning strategies, career planning, skill development, and academic experiences. Through iterative discussion and constant comparison, the researchers consolidated their memos into an initial code list, which was then refined through theoretical sensitization with existing literature on informal learning \citep{livingstone1999exploring, colley2003informality} and career development in higher education.

The resulting codebook, presented in Table~\ref{tab:codebook}, contains 7 categories and 25 codes, organized in a three-level hierarchical structure: category, code, and keywords. The revised structure separates disciplinary conditions from supply-side tensions, distinguishes career pathways from career orientations, and differentiates target capabilities from the learning pathways through which students seek to acquire them. Each code is operationalized through a set of characteristic keywords and phrases that serve as matching indicators. For example, within the career-pathway category, a code capturing graduate study intentions is associated with keywords such as ``postgraduate entrance exam,'' ``master's program,'' and ``entering higher schools.''

\begin{table*}[h]
\centering
\caption{Full Coding Framework: Categories, Codes, Coverage Rates, and Representative Examples}
\label{tab:codebook}
\small
\begin{tabular}{p{2cm} p{4cm} p{0.8cm} p{7.5cm}}
\hline
\textbf{Category} & \textbf{Code} & \textbf{Cov.} & \textbf{Representative Example (translated)} \\
\hline
\multirow{6}{2cm}{A: Major Characteristics}
  & Interdisciplinary character   &  2.7\% & ``This major crosses into EE, math, and management---it is hard to master all.'' \\
  & Practice orientation          & 18.6\% & ``Theory without practice is useless in this field.'' \\
  & Math/physics foundation       & 21.9\% & ``Calculus, linear algebra, and probability are the hardest gates---do not skip them.'' \\
  & Broad but not deep curriculum &  7.7\% & ``We cover everything superficially---depth is your own responsibility.'' \\
  & Outdated curriculum content   &  3.7\% & ``The textbooks teach frameworks nobody uses in industry anymore.'' \\
  & High learning difficulty      & 17.0\% & ``This major is genuinely hard---the dropout rate is high for a reason.'' \\
\hline
\multirow{2}{2cm}{B: Supply Tensions}
  & Weak teaching support         &  2.5\% & ``Some professors have not worked in industry for decades.'' \\
  & Formal course dissatisfaction &  4.6\% & ``Most of what I needed for my job I learned outside the classroom.'' \\
\hline
\multirow{3}{2cm}{C: Career Pathway}
  & Graduate study pathway        & 48.1\% & ``If you want to stay in research, postgraduate study is almost essential.'' \\
  & Public sector pathway         & 12.1\% & ``Civil service exams are a valid option---GPA and political theory matter there.'' \\
  & Enterprise employment pathway & 48.3\% & ``Internships at top tech firms are the real filter for industry jobs.'' \\
\hline
\multirow{4}{2cm}{D: Career Orientation}
  & Early career planning         & 17.4\% & ``Start planning your direction from day one---don't drift through freshman year.'' \\
  & Interest/fit assessment       & 29.9\% & ``Don't choose this major if you don't genuinely enjoy coding.'' \\
  & Major transfer/exit decision  & 11.9\% & ``If you truly hate engineering, transfer early---don't wait until year three.'' \\
  & Phased task consciousness     & 32.0\% & ``Year 1: foundations. Year 2: algorithms. Year 3: projects. Year 4: applications.'' \\
\hline
\multirow{4}{2cm}{E: Capability Building}
  & Programming \& algorithm core & 29.1\% & ``LeetCode is non-negotiable---start medium difficulty by sophomore year.'' \\
  & Self-directed learning        & 26.4\% & ``The ability to teach yourself is the most important skill in this major.'' \\
  & Complex problem solving       &  1.6\% & ``The ability to break down complex problems is what separates good engineers.'' \\
  & Communication \& teamwork     &  9.5\% & ``Soft skills matter more than people admit---practice presenting your work.'' \\
\hline
\multirow{4}{2cm}{F: Learning Pathway}
  & Online learning               & 13.2\% & ``Bilibili and Coursera cover the practical skills our courses skip.'' \\
  & Project-based practice        &  6.1\% & ``Build real projects, not toy demos. Open-source contributions count.'' \\
  & Competition training     & 28.3\% & ``ACM-ICPC training pays off for both graduate applications and job offers.'' \\
  & Internship experience         & 17.8\% & ``Get an internship by end of junior year---timing is everything.'' \\
\hline
\multirow{2}{2cm}{G: Achievement Recognition}
  & Credential/qualification conversion   &  6.0\% & ``Competition awards count toward extra credit and postgraduate bonus points.'' \\
  & Institutional recognition             &  6.0\% & ``Our department now counts certified open-source contributions for credit.'' \\
\hline
\end{tabular}
\end{table*}

To execute the coding at scale, a Python script was developed to perform keyword-based matching across the full corpus of 3,607 posts. The script scans each post for the presence of keywords associated with each code and produces a binary document-code matrix, where a value of 1 indicates that a given code was detected in a given post and 0 indicates absence. This approach enables systematic, reproducible coding while maintaining the semantic grounding established through human codebook construction.

Validation was conducted on a stratified random sample of 150 posts, comprising 100 posts that received at least one code assignment and 50 posts that remained uncoded. Two researchers independently reviewed these posts, comparing the automated coding output against their manual judgments. Inter-rater reliability was assessed using Cohen's $\kappa$ computed at the code level: each post--code pair (150 posts $\times$ 25 codes = 3,750 binary decisions) was treated as an independent observation. The overall $\kappa$ between the two human raters was 0.775, indicating substantial agreement \citep{landis1977measurement}. Agreement between the automated system and each rater was similarly high ($\kappa = 0.837$ and $\kappa = 0.817$, respectively). Per-code $\kappa$ values are reported in Table~\ref{tab:irr}; the majority of codes exceeded $\kappa = 0.70$, with lower values observed for codes whose boundaries are inherently more interpretive (e.g., \textit{perceived learning difficulty}, $\kappa = 0.359$; \textit{formal course dissatisfaction}, $\kappa = 0.280$). Discrepancies were analyzed to identify sources of error---primarily keyword omissions---and the codebook was revised accordingly. The revised script was then re-executed on the full corpus to produce the final coded dataset.

\begin{table*}[h]
\centering
\caption{Per-code Cohen's $\kappa$ between the two human raters (R1 vs.\ R2) and between the automated system and each rater (Auto vs.\ R1; Auto vs.\ R2), computed on the 150-post validation sample ($n = 3{,}750$ binary decisions)}
\begin{tabular}{llrccc}
\toprule
\textbf{Category} & \textbf{Code} & \textbf{\textit{n}} & $\boldsymbol{\kappa}$\textbf{(R1 vs.\ R2)} & $\boldsymbol{\kappa}$\textbf{(Auto vs.\ R1)} & $\boldsymbol{\kappa}$\textbf{(Auto vs.\ R2)} \\
\midrule
\multirow{6}{*}{A: Major Characteristics}
 & Interdisciplinary character   &  2 & 1.000 & 1.000 & 1.000 \\
 & Practice orientation          & 13 & 0.604 & 0.832 & 0.665 \\
 & Math/physics foundation       & 25 & 0.817 & 0.897 & 0.927 \\
  & Broad but not deep curriculum &  8 & 0.612 & 0.868 & 0.612 \\
  & Outdated curriculum content   &  7 & 0.650 & 0.920 & 0.736 \\
  & High learning difficulty      & 19 & 0.359 & 0.446 & 0.865 \\
\midrule
\multirow{2}{*}{B: Supply Tensions}
 & Weak teaching support         &  3 & 0.490 & 0.797 & 0.658 \\
 & Formal course dissatisfaction &  3 & 0.280 & 0.487 & 0.331 \\
\midrule
\multirow{3}{*}{C: Career Pathway}
 & Graduate study pathway        & 41 & 0.844 & 0.883 & 0.827 \\
 & Public sector pathway         &  7 & 0.764 & 1.000 & 0.764 \\
 & Enterprise employment pathway & 45 & 0.816 & 0.921 & 0.892 \\
\midrule
\multirow{4}{*}{D: Career Orientation}
 & Early career planning         &  9 & 0.612 & 0.521 & 0.407 \\
 & Interest--fit assessment      & 30 & 0.848 & 0.892 & 0.958 \\
 & Major transfer/exit decision  & 16 & 0.964 & 0.964 & 0.930 \\
 & Phased task consciousness     & 31 & 0.980 & 1.000 & 0.980 \\
\midrule
\multirow{4}{*}{E: Capability Building}
 & Programming \& algorithm core & 23 & 0.787 & 0.851 & 0.782 \\
 & Self-directed learning        & 28 & 0.814 & 0.656 & 0.683 \\
 & Complex problem solving       &  1 & 0.495 & 1.000 & 0.495 \\
 & Communication \& teamwork     &  7 & 0.528 & 0.759 & 0.827 \\
\midrule
\multirow{4}{*}{F: Learning Pathway}
 & Online learning               & 16 & 0.693 & 0.579 & 0.684 \\
 & Project-based practice        & 22 & 0.781 & 0.893 & 0.878 \\
 & Competition training     & 33 & 0.922 & 0.902 & 0.942 \\
 & Internship experience         & 19 & 1.000 & 1.000 & 1.000 \\
\midrule
\multirow{2}{*}{G: Achievement Recognition}
 & Credential/qualification conversion &  6 & 0.487 & 0.429 & 0.906 \\
 & Institutional recognition         &  7 & 0.324 & 0.598 & 0.582 \\
\midrule
\multicolumn{3}{l}{\textit{Mean}} & 0.699 & 0.804 & 0.773 \\
\multicolumn{3}{l}{\textit{Overall (all codes pooled)}} & 0.775 & 0.837 & 0.817 \\
\bottomrule
\end{tabular}
\label{tab:irr}
\end{table*}

To assess the empirical coverage of the codebook, we implemented a cumulative code-discovery analysis. The full set of 3,373 coded documents (at least one code) was randomly shuffled, and the corpus was traversed sequentially, recording the cumulative number of distinct codes observed after each document. Under this procedure, all 25 codes were first encountered within the first 98 documents (2.9\% of the corpus). To rule out dependence on a single random ordering, we repeated this procedure across 100 independent random permutations: the mean coverage point was 93 documents (2.8\%), with a 95\% bootstrap confidence interval of [36, 208] documents ([1.1\%, 6.2\%]). In every permutation, the codebook reached full coverage before 10\% of the corpus had been processed. These results indicate that the 25-code framework achieved comprehensive empirical coverage: all codes were represented in a small fraction of the corpus, suggesting that the codebook captures the full range of conceptually distinct themes present in the data. While this analysis does not guarantee theoretical saturation in the strict grounded-theory sense, the early coverage point supports the comprehensiveness of the coding framework within the boundaries of the current corpus.

\subsubsection{Phase 2: Axial Coding}

The axial coding phase examines relationships among the 25 codes by quantifying their document-level co-occurrence patterns. The unit of analysis is the individual post: two codes are considered co-occurring when both are present within the same document. Three complementary metrics are computed for every possible code pair, as summarized in Table~\ref{tab:metrics}.

\begin{table*}[h]
\centering
\caption{Co-occurrence metrics used in axial coding}
\begin{tabular}{lll}
\toprule
\textbf{Metric} & \textbf{Formula} & \textbf{Interpretation} \\
\midrule
Co-occurrence frequency & $f_{AB} = |A \cap B|$ & Absolute association count \\
Jaccard coefficient & $J_{AB} = \dfrac{|A \cap B|}{|A \cup B|}$ & Normalized overlap \\
Lift & $\text{lift}_{AB} = \dfrac{P(A,B)}{P(A) \cdot P(B)}$ & Above-independence association \\
PMI & $\text{PMI}_{AB} = \log_2(\text{lift}_{AB})$ & Log-scaled association strength \\
\bottomrule
\end{tabular}
\label{tab:metrics}
\end{table*}

Co-occurrence frequency provides a raw count of how often two codes appear together, capturing the absolute magnitude of their association. The Jaccard coefficient normalizes this count by the union of the two code sets, yielding a proportion that controls for base-rate differences across codes. Lift quantifies the degree to which two codes co-occur more frequently than would be expected under statistical independence; a lift value greater than 1.0 indicates positive association, while values below 1.0 indicate negative association. Pointwise mutual information (PMI), computed as $\log_2(\text{lift})$, provides a log-scaled representation of the same association that is more interpretable for comparing pairs with very different base-rate frequencies; it is reported in Appendix~\ref{app:axial_pairs} alongside lift and Jaccard for transparency.

To identify substantively meaningful code pairs for theoretical interpretation, we apply a dual threshold: a code pair is classified as strongly associated only if its lift exceeds 1.2 \emph{and} its co-occurrence frequency exceeds 200. The lift threshold ensures that the association is meaningfully above chance, while the frequency threshold ensures that it is grounded in a sufficient number of empirical observations. Applying these criteria to the full 298-pair matrix yields 42 qualifying code pairs. The Results section interprets these qualifying associations through Figure~\ref{fig:portfolio_networks} and the accompanying portfolio narratives. Complete axial coding results are provided in Appendix~\ref{app:axial_pairs}.

\subsubsection{Phase 3: Selective Coding}

Selective coding identifies the core category around which the emergent theory is organized. Following the CGT principle of combining quantitative evidence with interpretive reasoning, we compute a centrality score for each code that integrates two dimensions: the code's standalone prevalence and its relational embeddedness within the co-occurrence network. The centrality score is defined as:

\begin{equation}
C_i = 0.6 \times \hat{f}_i + 0.4 \times \widehat{wd}_i
\label{eq:centrality}
\end{equation}

\noindent where $\hat{f}_i$ denotes the min-max normalized document frequency of code $i$ (i.e., the proportion of posts in which the code appears, rescaled to the $[0,1]$ interval), and $\widehat{wd}_i$ denotes the min-max normalized weighted degree of code $i$ in the co-occurrence network, calculated as the sum of co-occurrence frequencies for all edges connecting code $i$ to codes with which it co-occurs. The weighting scheme assigns greater importance to document frequency (0.6) than to network centrality (0.4), reflecting the GT principle that a core category should be empirically pervasive, not merely well-connected.

The selective coding analysis reveals two codes with comparably high centrality scores: \emph{graduate study pathway} ($C = 0.997$) and \emph{enterprise employment pathway} ($C = 0.990$). These two codes appear in nearly identical shares of posts (48.1\% and 48.3\%, respectively) and maintain strong co-occurrence relationships with codes across multiple categories, confirming their joint integrative role in organizing the discourse. Complete selective coding results, including centrality scores for top 20 codes, are provided in Appendix~\ref{app:selective_coding}. 

Building on these dual core codes, the research team derives a candidate mechanism chain by examining how categories from the revised framework relate to the career pathway codes and to one another: \emph{major characteristics} shape \emph{career orientation}, which informs \emph{career pathway} selection (graduate study, enterprise employment, or public sector), motivates \emph{capability building}, and culminates in forms of \emph{achievement recognition}. This mechanism chain accommodates the three distinct portfolio configurations identified in the axial coding phase, each centered on a different career pathway orientation. The resulting theoretical framework is presented in the Results section.

\section{Results}\label{sec:results}

\subsection{Overview: Career-Oriented Discourse as Dominant Theme}

Open coding of the 3,607 valid posts yielded 3,373 posts assigned to at least one code, producing an overall coding coverage of 93.5\%. The seven emergent categories and their coverage rates are presented in Table~\ref{tab:coverage}.

\begin{table}[h]
\centering
\caption{Category coverage across 3,373 coded posts}
\begin{tabular}{lcc}
\toprule
\textbf{Category} & \textbf{Documents (\textit{n})} & \textbf{Coverage (\%)} \\
\midrule
Career pathway & 2,347 & 69.6 \\
Career orientation & 2,015 & 59.7 \\
Major characteristics & 1,686 & 50.0 \\
Capability building & 1,619 & 48.0 \\
Learning pathway & 1,582 & 46.9 \\
Achievement recognition & 375 & 11.1 \\
Supply tension & 225 & 6.7 \\
\bottomrule
\end{tabular}
\label{tab:coverage}
\end{table}

The distribution shows that career-oriented discourse occupies the center of peer advice. Career pathway was the most prevalent category (69.6\%), followed by career orientation (59.7\%). In other words, posts most often addressed not only \emph{where} students were headed (e.g., graduate study, enterprise employment, public-sector examinations) but also \emph{how} they should orient themselves toward those pathways (e.g., staged planning, interest-fit assessment, early decision making). Major characteristics (50.0\%), capability building (48.0\%), and learning pathway (46.9\%) formed a second tier, indicating that students typically discussed disciplinary conditions, target capabilities, and concrete acquisition routes as elaborations of career-oriented planning.

Achievement recognition (11.1\%) and supply tension (6.7\%) occupied the narrowest strata. Their lower prevalence should not be interpreted as analytical marginality. Instead, these categories appeared as more specialized moments in the discourse: students referred to achievement recognition when discussing how competitions, certificates, or policy rules convert activities into formal value, and to supply tension when explicitly diagnosing shortcomings in curricular or instructional provision.

At the individual-code level, the four highest-centrality codes were concentrated in the two career-related categories: graduate study pathway (48.1\%, centrality = 0.997), enterprise employment pathway (48.3\%, 0.990), phased task consciousness (32.0\%, 0.713), and interest/fit assessment (29.9\%, 0.637). The next highest-ranked codes were programming \& algorithm core (29.1\%, 0.637), competition training (28.3\%, 0.623), and self-directed learning (26.4\%, 0.581). Among major-characteristic codes, math/physics foundation (21.9\%, 0.485) ranked highest. Taken together, these patterns indicate that peer advice discourse is organized around career endpoints and career orientation, while specific capabilities, learning routes, and disciplinary constraints provide the substantive content through which those career narratives are operationalized. Centrality scores for the top 20 codes are reported in Appendix~\ref{app:selective_coding}.

\subsection{Three Career-Directed Digital Informal Learning Portfolios}

Selective coding identified two career pathway codes—graduate study pathway and enterprise employment pathway—as having the highest centrality scores, indicating their integrative role in organizing the discourse. Building on this finding, axial coding examined how these and other codes co-occurred to form coherent portfolio configurations.

Applying the dual threshold (lift $> 1.2$ and co-occurrence frequency $> 200$) to the full 298-pair matrix yielded 42 qualifying associations. These associations reveal that career-oriented discourse is not monolithic but differentiated into three pathway-centered portfolio types: one organized around graduate study preparation, one around enterprise employment preparation, and one characterized by dual-track hedging across multiple pathways. The three portfolios are interpreted by tracing how codes from multiple categories cluster around each career pathway anchor. Complete axial coding results are provided in Appendix~\ref{app:axial_pairs}.

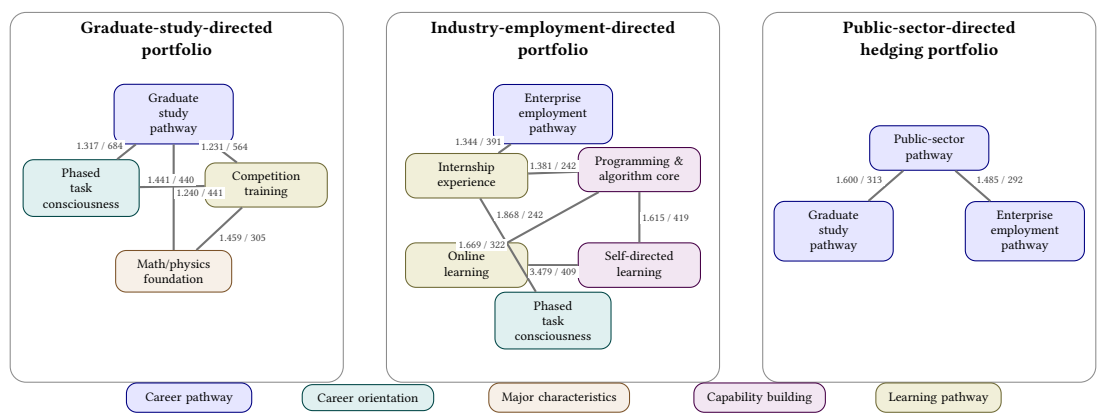
\begin{figure*}[t]
\centering
\begin{tikzpicture}[
  x=1cm, y=1cm, scale=0.82, transform shape,
  portfolio/.style = {
    rectangle, draw=black!45, rounded corners=6pt,
    minimum width=5.4cm, minimum height=6.0cm
  },
  titlebox/.style = {
    font=\small\bfseries, align=center
  },
  cpath/.style = {
    rectangle, rounded corners=4pt, draw=blue!50!black, fill=blue!10,
    align=center, font=\scriptsize, inner sep=4pt, text width=1.65cm
  },
  corient/.style = {
    rectangle, rounded corners=4pt, draw=teal!55!black, fill=teal!12,
    align=center, font=\scriptsize, inner sep=4pt, text width=1.6cm
  },
  cmajor/.style = {
    rectangle, rounded corners=4pt, draw=brown!65!black, fill=brown!12,
    align=center, font=\scriptsize, inner sep=4pt, text width=1.6cm
  },
  cskill/.style = {
    rectangle, rounded corners=4pt, draw=violet!55!black, fill=violet!10,
    align=center, font=\scriptsize, inner sep=4pt, text width=1.7cm
  },
  clearn/.style = {
    rectangle, rounded corners=4pt, draw=olive!60!black, fill=olive!12,
    align=center, font=\scriptsize, inner sep=4pt, text width=1.7cm
  },
  edge/.style = {draw=black!55, thick},
  edgelabel/.style = {font=\tiny, fill=white, inner sep=0.9pt, text=black!75}
]

% Panel frames
\node[portfolio] (p1) at (0,0) {};
\node[portfolio] (p2) at (6.1,0) {};
\node[portfolio] (p3) at (12.2,0) {};

\node[titlebox] at (0,2.55) {Graduate-study-directed\\portfolio};
\node[titlebox] at (6.1,2.55) {Industry-employment-directed\\portfolio};
\node[titlebox] at (12.2,2.55) {Public-sector-directed\\hedging portfolio};

% Panel 1
\node[cpath]  (gsp) at (-0.05,1.35) {Graduate\\study\\pathway};
\node[corient](gst) at (-1.55,0.15) {Phased\\task\\consciousness};
\node[clearn] (gct) at (1.45,0.2) {Competition\\training};
\node[cmajor] (gmf) at (-0.05,-1.2) {Math/physics\\foundation};

\draw[edge] (gsp) -- (gst) node[pos=0.54, above left, edgelabel] {1.317 / 684};
\draw[edge] (gsp) -- (gct) node[pos=0.48, above, edgelabel] {1.231 / 564};
\draw[edge] (gsp) -- (gmf) node[pos=0.46, right, edgelabel] {1.240 / 441};
\draw[edge] (gct) -- (gst) node[pos=0.52, above, edgelabel] {1.441 / 440};
\draw[edge] (gct) -- (gmf) node[pos=0.56, below right, edgelabel] {1.459 / 305};

% Panel 2
\node[cpath]  (iep) at (6.1,1.35) {Enterprise\\employment\\pathway};
\node[clearn] (iin) at (4.7,0.35) {Internship\\experience};
\node[cskill] (ipc) at (7.5,0.45) {Programming \&\\algorithm core};
\node[cskill] (isl) at (7.5,-1.1) {Self-directed\\learning};
\node[clearn] (iol) at (4.7,-1.1) {Online\\learning};
\node[corient](its) at (6.1,-2.0) {Phased\\task\\consciousness};

\draw[edge] (iep) -- (iin) node[pos=0.52, above left, edgelabel] {1.344 / 391};
\draw[edge] (iin) -- (its) node[pos=0.5, left, edgelabel] {1.669 / 322};
\draw[edge] (iin) -- (ipc) node[pos=0.5, above, edgelabel] {1.381 / 242};
\draw[edge] (iol) -- (isl) node[pos=0.5, below, edgelabel] {3.479 / 409};
\draw[edge] (iol) -- (ipc) node[pos=0.42, above left, edgelabel] {1.868 / 242};
\draw[edge] (ipc) -- (isl) node[pos=0.52, right, edgelabel] {1.615 / 419};

% Panel 3
\node[cpath]  (pps) at (12.2,0.8) {Public-sector\\pathway};
\node[cpath]  (pgs) at (10.65,-0.55) {Graduate\\study\\pathway};
\node[cpath]  (pee) at (13.75,-0.55) {Enterprise\\employment\\pathway};

\draw[edge] (pps) -- (pgs) node[pos=0.52, above left, edgelabel] {1.600 / 313};
\draw[edge] (pps) -- (pee) node[pos=0.52, above right, edgelabel] {1.485 / 292};

% Compact legend bar
\node[cpath,  minimum width=2.0cm, text width=1.75cm]  at (0.2,-3.25)  {Career pathway};
\node[corient,minimum width=2.1cm, text width=1.85cm]  at (3.1,-3.25)  {Career orientation};
\node[cmajor, minimum width=2.25cm, text width=2.0cm] at (6.2,-3.25)  {Major characteristics};
\node[cskill, minimum width=2.1cm, text width=1.85cm]  at (9.45,-3.25) {Capability building};
\node[clearn, minimum width=2.0cm, text width=1.75cm]  at (12.35,-3.25) {Learning pathway};

\end{tikzpicture}
\caption{Three career-directed \textit{digital informal learning portfolios} derived from axial coding and selective coding. Nodes represent the most consequential codes used in the portfolio interpretations; edge labels report lift and co-occurrence frequency (\textit{n}). In Panel 2, the online learning--self-directed learning--programming triad has no qualifying direct association with the enterprise employment pathway code; it is included because its constituent codes qualify through internship experience as an intermediary (internship $\leftrightarrow$ programming: lift = 1.381; internship $\leftrightarrow$ self-directed learning: lift = 1.314). The public-sector portfolio is visualized as a hedging structure across career endpoints rather than a dense capability bundle.}
\label{fig:portfolio_networks}
\Description{Three side-by-side network diagrams summarize the graduate-study, industry-employment, and public-sector portfolios. Colors distinguish career pathway, career orientation, major characteristics, capability building, and learning pathway nodes. Edge labels show lift and co-occurrence frequency. The first two panels show pathway-centered learning bundles, and the third shows a hedging structure linking public-sector, graduate-study, and enterprise-employment pathways.}
\end{figure*}

To make these configurations visually tractable, Figure~\ref{fig:portfolio_networks} renders each portfolio as a compact network of the most consequential codes and co-occurrence relations. The first two subfigures highlight pathway-centered learning architectures anchored by the two high-centrality career codes, whereas the third highlights a hedging structure that maintains readiness across multiple career endpoints.

\subsubsection{Portfolio 1: Graduate-Study-Directed Portfolio}

The graduate-study-directed portfolio was organized around the graduate study pathway code (48.1\% coverage, centrality = 0.997) and was distinguished by its tight integration of competition training, mathematical foundations, and developmental sequencing.

A key cross-category association linked competition training and math/physics foundation (lift = 1.459, \textit{n} = 305), indicating that posts discussing mathematical and physical foundations were substantially more likely to co-occur with competition training than would be expected under independence. Competition training also exhibited strong co-occurrence with the graduate study pathway itself (lift = 1.231, \textit{n} = 564), and the graduate study pathway was directly associated with math/physics foundation (lift = 1.240, \textit{n} = 441). Together, these patterns suggest that competitions and foundational coursework were jointly understood as preparation resources for postgraduate progression.

Within this portfolio, phased task consciousness played a structuring role. The co-occurrence between graduate study pathway and phased task consciousness was substantial (lift = 1.317, \textit{n} = 684), and competition training similarly co-occurred with phased task consciousness (lift = 1.441, \textit{n} = 440). These associations indicate that students oriented toward graduate study did not merely accumulate activities but organized them into developmentally sequenced stages. The graduate study pathway was also positively associated with high learning difficulty (lift = 1.317, \textit{n} = 362), suggesting that progression-oriented advice was often framed against the backdrop of demanding disciplinary foundations.

The qualitative data corroborated this staged, foundation-first logic. As one contributor advised:

% Source: data/data.csv, id=1948, 重庆邮电大学, 计算机科学与技术
\begin{quote}
``If, like me, you are not especially strong at coding and are not planning to start a business or enter employment right away, you can begin preparing for postgraduate entrance exams in your junior year. Across your four undergraduate years, the key is not to fail courses, to master the basic professional knowledge, and at least to become proficient in one programming language.'' [translated by authors]
\end{quote}

Another post articulated the phased structure explicitly:

% Source: data/data.csv, id=557, 淮南师范学院, 电子信息工程
\begin{quote}
``Freshman year is mainly for foundational courses and general education, with your main effort going into calculus, linear algebra, probability, and C programming. Sophomore year is when you start learning microcontrollers and trying competitions. Junior year is mainly for competitions and accumulating field-related awards. Senior year is for either postgraduate entrance exams or job applications.'' [translated by authors]
\end{quote}

Taken together, the quantitative co-occurrence patterns and qualitative evidence revealed a portfolio in which mathematical foundations, competition training, and staged planning formed a unified preparation system. The informal learning activities were not ad hoc supplements to the formal curriculum but constituted a coherent, career-directed learning architecture aimed at maximizing competitiveness in graduate admissions.

\subsubsection{Portfolio 2: Industry-Employment-Directed Portfolio}

The industry-employment-directed portfolio centered on the enterprise employment pathway code (hereafter used interchangeably as ``industry employment''; 48.3\% coverage, centrality = 0.990) and was characterized most clearly by strategic internship accumulation embedded in a broader capability-building repertoire.

The strongest direct association between enterprise employment pathway and a learning-pathway code was internship experience (lift = 1.344, \textit{n} = 391), confirming internship as the most distinctive empirical anchor of the employment-directed portfolio. Around this anchor, the broader co-occurrence network included a tightly linked capability-building triad of online learning, self-directed learning, and programming \& algorithm core: online learning co-occurred strongly with self-directed learning (lift = 3.479, \textit{n} = 409), online learning also linked to programming \& algorithm core (lift = 1.868, \textit{n} = 242), and programming \& algorithm core co-occurred with self-directed learning (lift = 1.615, \textit{n} = 419). Importantly, these three codes have no qualifying direct association with the enterprise employment pathway code itself; they are connected to the employment-directed portfolio through internship experience as an intermediary---internship co-occurs with programming \& algorithm core (lift = 1.381, \textit{n} = 242) and with self-directed learning (lift = 1.314, \textit{n} = 209). The triad thus represents a general skill-building infrastructure that becomes employment-relevant through its link to strategically timed internship acquisition rather than through direct co-occurrence with the enterprise employment pathway.

Internship experience also co-occurred strongly with phased task consciousness (lift = 1.669, \textit{n} = 322), and with programming \& algorithm core (lift = 1.381, \textit{n} = 242). This pattern shows that internship was not treated as an isolated experience but as a temporally sequenced outcome of skill accumulation. Enterprise employment pathway additionally co-occurred with high learning difficulty (lift = 1.259, \textit{n} = 348). A comparable association exists between graduate study pathway and high learning difficulty (lift = 1.317, \textit{n} = 362), indicating that awareness of demanding preparation requirements is a cross-pathway feature of STEM peer advice rather than a marker specific to employment-oriented discourse.

The qualitative data further suggested a recurring sequence in which platform-mediated skill acquisition was translated into demonstrable outputs and then into targeted internship opportunities:

% Source: data/data.csv, id=1864, 黑龙江工程学院, 计算机科学与技术
\begin{quote}
``Use MOOCs, Bilibili, NetEase Open Courses, and Coursera to teach yourself programming; many university instructors simply read from slides, and some have never done projects themselves, so high-quality online courses can be more rewarding. Do more computing-related projects, whether entrepreneurial projects, lab projects, or GitHub projects, to build hands-on experience for future job hunting. And practice LeetCode, practice LeetCode, practice LeetCode---you can start in your sophomore year.'' [translated by authors]
\end{quote}

The emphasis on self-directed, platform-mediated learning was explicit:

% Source: data/data.csv, id=1882, 锦城学院, 计算机科学与技术
\begin{quote}
``From freshman year onward, do not just follow the university curriculum. Learn to search for information online, not only for study materials but also to use other people's experience to plan your university life as efficiently as possible. For employment, look at job postings, identify the skills they require, teach yourself those skills, and try to secure an internship by the summer after sophomore year.'' [translated by authors]
\end{quote}

This portfolio thus represented a coherent informal learning sequence in which self-directed digital study and programming-focused capability building prepared students for strategically timed internships, often supplemented by project work that could be mobilized in recruitment contexts. Compared with the graduate-study portfolio, the employment-directed type was less tightly tied to a single high-lift preparation bundle and more embedded in a broader, modular skill-building repertoire. The formal curriculum was not rejected outright, but it was often treated as a baseline rather than a sufficient pathway to employability.

\subsubsection{Portfolio 3: Public-Sector-Directed Portfolio and Dual-Track Hedging}

The public-sector-directed portfolio was the least prevalent of the three types, anchored by the public sector pathway code (12.1\% coverage, centrality = 0.275). However, its co-occurrence structure revealed a distinctive strategic logic that set it apart from the other portfolios.

The most striking finding was the exceptionally high inter-path co-occurrence between graduate study pathway and public sector pathway (lift = 1.600, \textit{n} = 313), the highest lift value between any two career-pathway codes. Enterprise employment pathway and public sector pathway also co-occurred at an elevated rate (lift = 1.485, \textit{n} = 292). These associations indicate that public-sector-oriented students did not pursue a singular pathway but engaged in systematic dual-track hedging, maintaining simultaneous readiness for graduate study, enterprise employment, and public-sector entry.

This hedging strategy had direct implications for portfolio composition. Unlike the specialized portfolios described above, the public-sector portfolio produced no qualifying co-occurrence associations with specific capability-building or learning-pathway codes. Two factors contribute to this pattern. Substantively, public-sector examinations and civil service recruitment reward GPA, broad knowledge, and institutional credentials, reducing incentives for deep investment in competition training, project building, or platform-based skill acquisition. Methodologically, the public-sector pathway code appears in only 407 documents (12.1\% of coded posts), substantially fewer than the graduate-study (48.1\%) and enterprise employment (48.3\%) codes; the $n>200$ co-occurrence threshold therefore requires more than 49\% of public-sector posts to contain a partner code before a pair qualifies, a considerably higher bar than for larger categories. Several public-sector associations show elevated lift values---including public-sector pathway with institutional recognition (lift = 2.08) and with transfer/exit decision (lift = 1.93)---but fall below the frequency threshold. The portfolio's apparent generalism should therefore be interpreted with this threshold asymmetry in mind, rather than taken as definitive evidence of undifferentiated learning behavior.

The qualitative data captured this hedging orientation concisely:

% Source: data/data.csv, id=3436, 安徽财经大学, 土地资源管理
\begin{quote}
``Plan early. For our major, the best employment route is the civil service. If you plan to take postgraduate entrance exams, you should also prepare for the civil service exam at the same time and keep both options open.'' [translated by authors]
\end{quote}

The public-sector portfolio thus represented a risk-diversification strategy in which the formal curriculum retained greater salience than in the other two portfolios. Rather than constructing a deeply specialized informal learning architecture, these students maintained a generalist profile that preserved optionality across multiple career endpoints.

Across all three portfolio types, a common temporal logic recurs: community advice consistently urges students to orient career goals and begin purposeful learning investment well before formal curricula direct them to do so. This early-career-organizing pattern---visible in the high prevalence of phased task consciousness (32.0\%) and early career planning (17.4\%) codes and their strong associations with career pathway codes---is theorized in Section~\ref{sec:discussion} as \textit{career front-loading}.

\subsection{The Peer Advice Community as Distributed Informal Curriculum}

Beyond the three portfolio types, the analysis yielded a second-order finding concerning the nature of the peer advice community itself. The 3,607 posts in the corpus did not merely contain isolated fragments of personal experience; collectively, they constituted a structured, pathway-differentiated body of knowledge about career-learning navigation in STEM higher education. The prominence of career pathway, career orientation, and staged-task discourse indicates that community knowledge was organized not simply by major identity, but by pathway choice, developmental timing, and the practical means of capability acquisition. In aggregate, this body of knowledge functioned as a distributed informal curriculum.

The peer advice community exhibited the core features of a community of practice \citep{wenger1998communities}: a shared domain, a cross-cohort community, and a shared practice of articulating and refining portfolio-construction strategies. Junior students entering the community encountered an established repertoire of portfolio blueprints, sequencing recommendations, and platform evaluations transmitted by seniors and alumni.

A critical characteristic of this distributed curriculum was its systematic evaluation of external digital platforms. Community contributors frequently rated platforms such as LeetCode, Bilibili, Coursera, and GitHub as superior to formal course content for the development of practically relevant skills. This evaluative function meant that the community did not merely supplement the formal curriculum; it actively redirected students' learning investments toward informal digital resources that were perceived as more closely aligned with career demands. As one senior student reflected:

% Source: data/data.csv, id=2721, 广西科技大学, 软件工程
\begin{quote}
``Ask senior students; they often have review materials and resources you would not think of on your own. But more importantly, in this major you need to code a lot, teach yourself, and learn ahead of the university's teaching schedule. If you simply follow the school's pace, you will find it very hard later. You need to identify the technical routes required by industry and start learning along those routes early.'' [translated by authors]
\end{quote}

The early career planning code (17.4\% coverage) and its co-occurrence with graduate study pathway (lift = 1.258, \textit{n} = 355), together with the strong association between early career planning and phased task consciousness (lift = 1.478, \textit{n} = 278), further underscored the community's curricular function. Planning-oriented discourse was systematically tied to pathway choice and temporal sequencing, indicating that the community served not only as an information repository but as an active socializing agent, instilling in newcomers the imperative of pathway-specific portfolio construction.

This distributed curriculum finding frames the theoretical discussion in Section~\ref{sec:discussion_distributed_curriculum}, where we examine how peer advice communities mediate the relationship between formal higher education institutions and the informal digital learning ecology that increasingly shapes STEM students' career preparation.

\section{Discussion}\label{sec:discussion}

\subsection{Career Goals as the Organizing Principle of Digital Informal Learning}

The finding that career pathway and career orientation are the two most prevalent categories challenges influential accounts of informal learning as primarily interest-driven or curiosity-motivated \citep{livingstone1999exploring, marsick2001informal}. In the internet-mediated STEM higher education context studied here, informal learning is instead organized largely around anticipated career outcomes and the temporal orientation required to pursue them. This pattern resonates strongly with SCCT, which argues that outcome expectations help organize career-relevant behavior \citep{lent1994toward,lent2002social}; our findings extend that logic to portfolio-level learning behavior spanning platforms, activities, and timeframes. The concept of \textit{career front-loading} captures this pattern. The discourse suggests that students orient their career judgments and frame their learning investments early---often well before completing formal curricula---in ways that challenge the conventional sequence in which disciplinary learning is assumed to precede career decision-making.

This finding extends SCCT in a consequential way. Rather than functioning only as triggers for discrete career choices, career cognitions in our data operate as persistent portfolio-organizing forces that shape which skills students prioritize, which platforms they invest time in, and which credentials they pursue over multiple years. The portfolio-level coherence observed across career orientations further substantiates this claim. Students are not simply choosing isolated activities; they are assembling integrated, pathway-specific learning architectures across internet-based platforms and peer communities. Informal learning, in this sense, is not the opposite of formal learning but a strategic extension of it within a broader higher education learning ecology.

\subsection{The Peer Advice Community as Digital Informal Curriculum}\label{sec:discussion_distributed_curriculum}

The finding that the online peer advice community functions as a structured knowledge system extends CoP theory \citep{lave1991situated} into the domain of internet-enabled higher education peer learning \citep{chen_new_2025}. Rather than merely exchanging tips, participants collectively produce and refine pathway-specific roadmaps for what to learn, when to learn it, where to learn it online, and how those learning investments convert into career capital.

A key distinction from traditional CoP formulations, however, lies in the \textit{guidance infrastructure function} this community performs. Classical CoP theory describes communities that complement institutional learning through situated practice and legitimate peripheral participation \citep{lave1991situated}. The community studied here goes further: it provides internet-mediated career-learning integration guidance that formal higher education institutions only partially provide. When universities offer disciplinary knowledge but limited guidance on how that knowledge maps onto differentiated career pathways and online learning opportunities, the peer community fills the gap with detailed, experience-based portfolio advice. This positions the community not simply as a supplement to institutional education but as a distributed knowledge infrastructure within the higher education internet ecology \citep{dron_pedagogical_2022}. In this context, the internet functions not merely as a repository of resources but as an organized layer of learning guidance, absorbing advisory responsibilities that traditional academic structures have not fully adapted to fulfill.

\citet{bourdieu1986forms} capital theory provides an additional analytical lens. Although achievement-recognition codes appeared less frequently than pathway and orientation codes, they reveal that the community implicitly encodes conversion rules---specifying how informal capital (competition awards, project work, internships, and certificates) translates into formal credentials and career advantages. Community discourse does not merely recommend activities; it also clarifies which achievements are institutionally legible and where informal effort can be converted into recognizable value. In this sense, the community operates as a ``capital conversion guide'' rather than merely a peer support forum. Students who participate gain not only specific knowledge but also an understanding of the rules governing capital conversion across institutional boundaries.

The implication for higher education institutions is consequential. The question is not whether to tolerate or encourage extracurricular internet-enabled learning---it is already normative among STEM students and deeply embedded in their developmental strategies. The more pressing question is whether institutions will recognize the portfolio infrastructure that students are already constructing and take steps to reduce the friction of credential conversion and guidance access. Failing to do so does not prevent portfolio construction; it merely increases the transaction costs borne by students and widens the advantage held by those with superior access to community knowledge.

\subsection{Methodological Contribution and Limitations}

This study demonstrates the productive transfer of CGT \citep{nelson2020computational} from its original sociological use case to higher education online community data. The main methodological contribution lies in showing that CGT can scale interpretive analysis to a corpus of more than 3,600 posts while preserving theoretically meaningful categories, transparent co-occurrence criteria, and an explicitly justified core-category selection process. This positions CGT as a useful middle ground between small-N qualitative studies and large-N unsupervised text analysis.

Several limitations warrant acknowledgment. First, the keyword-based coding approach, while scalable, is less sensitive to figurative language, negation, and sarcasm than human interpretation. This limitation is inherent to the computational layer of CGT and partially mitigated by the human interpretive layer, but it cannot be fully eliminated without more sophisticated natural language processing techniques. Second, the sample reflects platform selection bias. Students who compose detailed major advice posts in online communities may possess stronger reflective capacity, writing skills, and career awareness than the general STEM student population. The findings may therefore overrepresent ``successful navigators''---students who have already developed effective portfolio strategies---while underrepresenting those who struggle with career-learning integration. Third, the cross-sectional design captures portfolio descriptions at a single point in time and cannot trace how portfolios are constructed, revised, or abandoned over the undergraduate lifecycle. Students' retrospective accounts of their learning trajectories may be subject to post-hoc rationalization, presenting greater coherence than the actual decision-making process entailed. Fourth, the data derive from a single national context. China-specific labor market structures---including intense competition for graduate study admissions, the prominence of civil service examinations as a career pathway, and particular industry hiring practices---likely shape the portfolio types identified in this study. Whether the same three portfolio orientations emerge in higher education systems with different career structures, labor market conditions, and institutional arrangements remains an open empirical question.
\section{Conclusion}\label{sec:conclusion}

\subsection{Summary of Main Findings}

\textbf{Finding 1:} Career pathway and career orientation constitute the dominant organizing dimensions of digital informal learning among STEM students, and the discourse reflects a pattern of \textit{career front-loading} in which career judgments are framed as organizing learning investments early in the undergraduate lifecycle.

\textbf{Finding 2:} Three career-differentiated digital informal learning portfolios emerge from the data, showing that informal learning activities are assembled into coherent, pathway-specific configurations rather than pursued as isolated supplements.

\textbf{Finding 3:} The online peer advice community functions as a distributed informal curriculum that supplies pathway-specific guidance, staged planning knowledge, and capital-conversion knowledge within the broader internet ecology of higher education.

\subsection{Theoretical Contributions}

This study makes three primary theoretical contributions to the literature on internet-enabled learning and career development in higher education. First, it introduces the concept of the \textit{digital informal learning portfolio} as an analytic lens for understanding how STEM students combine and sequence multiple online learning activities in relation to anticipated career pathways. Second, it extends SCCT \citep{lent1994toward, lent2002social} from career choice into the organization of digital informal learning behavior, and introduces \textit{career front-loading} as a concept for understanding the early career-oriented framing of learning investments around anticipated career outcomes. Third, it conceptualizes the online peer advice community as a distributed informal curriculum that collectively produces and transmits pathway-specific guidance about valued skills, appropriate timing, and capital conversion within the internet ecology of higher education.

\subsection{Practical Implications}

The findings carry three practical implications. First, for curriculum designers and academic advisors, career front-loading suggests that career-learning integration should not be deferred to the final year of undergraduate study. Institutions should introduce early-year career exploration modules, structured online guidance, and flexible curriculum pathways that allow students to align coursework with emerging career orientations, rather than assuming a fixed sequence of disciplinary preparation followed by career decision-making.

Second, for institutions more broadly, formal credential recognition for portfolio elements---including competition awards, open-source project contributions, and structured internship experiences---would reduce the structural friction that students currently navigate informally across internet-based learning environments. Establishing transparent conversion rules between informal learning achievements and institutional credit or credential recognition would lower transaction costs for all students and reduce the informational advantages currently held by those with superior access to peer community knowledge.

Third, for digital platform developers and higher education service providers, the multi-platform learning behavior documented in this study suggests demand for portfolio-level tools that help students plan and track cross-platform developmental sequences. Current platforms typically support individual learning activities in isolation; tools that integrate planning, guidance, and recognition across platforms would better serve the actual portfolio construction behavior that students already engage in.

\subsection{Future Research Directions}

Several directions merit future investigation. Cross-national studies should examine whether the same portfolio types emerge under different labor-market and institutional conditions. Longitudinal designs should trace how digital informal learning portfolios evolve over the undergraduate lifecycle and what outcomes they predict. More advanced natural language processing techniques may help address the keyword-sensitivity limitations of the current approach. Finally, future work should examine whether access to portfolio-construction resources and community knowledge is distributed equitably across student groups.

\appendix
\section{Axial Coding Results: All Qualifying Code Co-occurrence Pairs}
\label{app:axial_pairs}

Table~\ref{tab:axial_pairs} presents all 42 code co-occurrence pairs satisfying the dual threshold (lift $>1.2$ and $n>200$), sorted by lift in descending order. These pairs constitute the full empirical basis for the portfolio configurations identified in Section~\ref{sec:results}.

\begin{table*}[htbp]
\centering
\caption{Axial Coding Results: All 42 Qualifying Code Co-occurrence Pairs (lift $>1.2$, $n>200$)}
\label{tab:axial_pairs}
\footnotesize
\begin{tabular}{llllrrrr}
\toprule
\textbf{Code A} & \textbf{Cat.} & \textbf{Code B} & \textbf{Cat.} & \textbf{\textit{n}} & \textbf{Lift} & \textbf{Jaccard} & \textbf{PMI} \\
\midrule
Online learning                  & F & Self-directed learning          & E & 409 & 3.479 & 0.441 & 1.799 \\
Online learning                  & F & Programming \& algorithm core   & E & 242 & 1.868 & 0.204 & 0.901 \\
Internship experience            & F & Phased task consciousness       & D & 322 & 1.669 & 0.237 & 0.739 \\
Competition training             & F & Online learning                 & F & 204 & 1.623 & 0.171 & 0.698 \\
Programming \& algorithm core    & E & Self-directed learning          & E & 419 & 1.615 & 0.288 & 0.692 \\
Graduate study pathway           & C & Public sector pathway           & C & 313 & 1.600 & 0.183 & 0.678 \\
Competition training             & F & Practice orientation            & A & 278 & 1.564 & 0.213 & 0.646 \\
Practice orientation             & A & Programming \& algorithm core   & E & 284 & 1.551 & 0.214 & 0.633 \\
Enterprise employment pathway    & C & Public sector pathway           & C & 292 & 1.485 & 0.167 & 0.570 \\
Early career planning            & D & Phased task consciousness       & D & 278 & 1.478 & 0.200 & 0.563 \\
Online learning                  & F & Phased task consciousness       & D & 210 & 1.472 & 0.160 & 0.558 \\
Competition training             & F & Math/physics foundation         & A & 305 & 1.459 & 0.220 & 0.545 \\
Competition training             & F & Programming \& algorithm core   & E & 404 & 1.456 & 0.264 & 0.542 \\
Competition training             & F & Phased task consciousness       & D & 440 & 1.441 & 0.276 & 0.527 \\
High learning difficulty         & A & Phased task consciousness       & D & 260 & 1.418 & 0.187 & 0.504 \\
Internship experience            & F & Programming \& algorithm core   & E & 242 & 1.381 & 0.180 & 0.465 \\
Early career planning            & D & Interest/fit assessment         & D & 240 & 1.369 & 0.177 & 0.454 \\
Math/physics foundation          & A & Phased task consciousness       & D & 322 & 1.358 & 0.215 & 0.441 \\
Self-directed learning           & E & Phased task consciousness       & D & 385 & 1.348 & 0.243 & 0.431 \\
Enterprise employment pathway    & C & Internship experience           & F & 391 & 1.344 & 0.212 & 0.427 \\
Programming \& algorithm core    & E & Phased task consciousness       & D & 422 & 1.341 & 0.257 & 0.423 \\
Math/physics foundation          & A & Programming \& algorithm core   & E & 288 & 1.337 & 0.201 & 0.419 \\
Graduate study pathway           & C & High learning difficulty        & A & 362 & 1.317 & 0.198 & 0.397 \\
Graduate study pathway           & C & Phased task consciousness       & D & 684 & 1.317 & 0.339 & 0.397 \\
Competition training             & F & Self-directed learning          & E & 331 & 1.315 & 0.219 & 0.395 \\
Internship experience            & F & Self-directed learning          & E & 209 & 1.314 & 0.163 & 0.394 \\
Enterprise employment pathway    & C & Major transfer/exit decision    & D & 249 & 1.279 & 0.140 & 0.355 \\
Interest/fit assessment          & D & Practice orientation            & A & 240 & 1.278 & 0.172 & 0.354 \\
Practice orientation             & A & Self-directed learning          & E & 212 & 1.276 & 0.162 & 0.352 \\
Graduate study pathway           & C & Major transfer/exit decision    & D & 247 & 1.275 & 0.139 & 0.351 \\
Early career planning            & D & Programming \& algorithm core   & E & 217 & 1.270 & 0.161 & 0.345 \\
Competition training             & F & High learning difficulty        & A & 205 & 1.268 & 0.155 & 0.343 \\
Interest/fit assessment          & D & Math/physics foundation         & A & 280 & 1.267 & 0.191 & 0.342 \\
Interest/fit assessment          & D & Phased task consciousness       & D & 409 & 1.267 & 0.244 & 0.342 \\
Enterprise employment pathway    & C & High learning difficulty        & A & 348 & 1.259 & 0.188 & 0.332 \\
Early career planning            & D & Graduate study pathway          & C & 355 & 1.258 & 0.192 & 0.332 \\
Competition training             & F & Early career planning           & D & 208 & 1.254 & 0.156 & 0.327 \\
Competition training             & F & Internship experience           & F & 212 & 1.246 & 0.158 & 0.318 \\
High learning difficulty         & A & Interest/fit assessment         & D & 212 & 1.241 & 0.155 & 0.312 \\
Graduate study pathway           & C & Math/physics foundation         & A & 441 & 1.240 & 0.230 & 0.310 \\
Competition training             & F & Graduate study pathway          & C & 564 & 1.231 & 0.281 & 0.300 \\
Interest/fit assessment          & D & Internship experience           & F & 220 & 1.224 & 0.158 & 0.292 \\
\bottomrule
\end{tabular}
\smallskip\\
\footnotesize\textit{Note.} Category abbreviations match Table~\ref{tab:codebook}: A = Major Characteristics, C = Career Pathway, D = Career Orientation, E = Capability Building, F = Learning Pathway. \textit{n} = documents containing both codes; Lift = observed/expected co-occurrence ratio; Jaccard = $|A \cap B|/|A \cup B|$; PMI = $\log_2(\text{lift})$. Pairs are sorted by lift in descending order.
\end{table*}

\section{Selective Coding Results: Code Centrality Scores}
\label{app:selective_coding}

Table~\ref{tab:ranked_codes} presents the complete ranked list of codes from selective coding, ordered by centrality score. The centrality score reflects each code's structural importance in the co-occurrence network, calculated using weighted degree centrality normalized across all codes.

\begin{table*}[htbp]
\centering
\caption{Selective Coding Results: Ranked Codes by Centrality Score}
\label{tab:ranked_codes}
\footnotesize
\begin{tabular}{llrrrr}
\toprule
\textbf{Code Label} & \textbf{Category} & \textbf{Doc Count} & \textbf{Weighted Degree} & \textbf{Degree} & \textbf{Centrality} \\
\midrule
\textbf{Graduate study pathway} & Career Pathway & 1,621 & 7,187 & 24 & 0.997 \\
\textbf{Enterprise employment pathway} & Career Pathway & 1,630 & 7,002 & 24 & 0.990 \\
Phased task consciousness & Career Orientation & 1,081 & 5,665 & 24 & 0.713 \\
Interest/fit assessment & Career Orientation & 1,007 & 4,794 & 24 & 0.637 \\
Programming \& algorithm core & Capability Building & 982 & 4,959 & 24 & 0.637 \\
Competition training & Learning Pathway & 953 & 4,887 & 24 & 0.623 \\
Self-directed learning & Capability Building & 891 & 4,548 & 24 & 0.581 \\
Math/physics foundation & Major Characteristics & 740 & 3,814 & 24 & 0.485 \\
Practice orientation & Major Characteristics & 629 & 3,274 & 24 & 0.414 \\
Internship experience & Learning Pathway & 602 & 3,256 & 24 & 0.403 \\
Early career planning & Career Orientation & 587 & 3,108 & 24 & 0.389 \\
High learning difficulty & Major Characteristics & 572 & 3,078 & 24 & 0.382 \\
Online learning & Learning Pathway & 445 & 2,682 & 24 & 0.313 \\
\textbf{Public sector pathway} & Career Pathway & 407 & 2,242 & 24 & 0.275 \\
Major transfer/exit decision & Career Orientation & 403 & 1,957 & 24 & 0.257 \\
Communication \& teamwork & Capability Building & 319 & 1,867 & 24 & 0.221 \\
Broad but not deep curriculum & Major Characteristics & 259 & 1,318 & 24 & 0.169 \\
Credential/qualification conversion & Achievement Recognition & 201 & 1,356 & 24 & 0.149 \\
Institutional recognition & Achievement Recognition & 203 & 1,241 & 24 & 0.144 \\
Project-based practice & Learning Pathway & 205 & 1,196 & 24 & 0.142 \\
\bottomrule
\end{tabular}
\smallskip\\
\footnotesize\textit{Note.} Doc Count = number of documents containing the code; Weighted Degree = sum of co-occurrence frequencies; Degree = number of unique code co-occurrences; Centrality = normalized weighted degree centrality score.
\end{table*}

\begin{acks}
The authors thank the student community (http://kkdaxue.com/) members whose peer advice posts formed the empirical basis of this study. All data were collected and processed in accordance with institutional ethics guidelines. The research received no specific grant from any funding agency in the public, commercial, or not-for-profit sectors.
\end{acks}

\balance
\clearpage
\bibliographystyle{ACM-Reference-Format}
\bibliography{sample-base}

\end{document}